\documentclass[aip,jap,endfloats,preprint]{revtex4-1}
\usepackage{graphicx,amssymb}
\usepackage{dcolumn}
\begin{document}
\preprint{AIP}
\title{Comment on ``First-principles studies on magnetic properties of rocksalt structure $M$C ($M$=Ca,Sr, and Ba) under pressure''[Appl. Phys. Lett. 98,182501(2011)] }
\author{Wenxu Zhang}\email{xwzhang@uestc.edu.cn}
\affiliation{SKLETFID, University of Electronic Science and
Technology of China, Chengdu 610054, China}
\author{Zhida Song}
\affiliation{School of Microelectronics and Solid State Electronics,
University of Electronic Science and Technology of China, Chengdu
610054, China}
\author{ Wanli Zhang}
\affiliation{SKLETFID, University of Electronic Science and
Technology of China, Chengdu 610054, China}
\date{\today}
\maketitle
\par A recent Letter\cite{dong11} reported magnetic properties and
electronic structures of $M$C ($M$=Ca, Sr, and Ba) under pressure.
As shown in Fig.1 in the Letter\cite{dong11}, there are
discontinuous jumps in the magnetic moment when the lattice
constants are decreased. There is also intermediate spin magnetic
moment in a quite wide range of the lattice constant in SrC and CaC.
However, these results come from the inaccuracy of the calculation:
the non-convergence of the spin moment with respect to the k-point
mesh in the Brillouin zone.
\par
\begin{figure}
\includegraphics[width=0.5\textwidth]{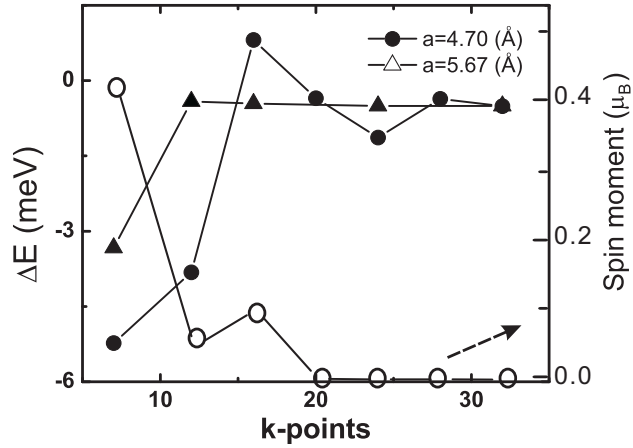}
\caption{The ground state energy at lattice constant of 4.70 \AA\
(filled circles) and 5.67 \AA\ (filled triangles) and the spin
magnetic moment (hollow circles) at lattice constant of 4.70 \AA\
after self-consistent calculations vs. k-points for SrC. The
abscissa are the k-point in one direction of the Brillouin zone. The
total number of the k-points is the cubic of the
abscissa.}\label{fig:ek}
\end{figure}

 With the input parameters provided by the authors, by using the same code we obtained
that the spin moment are 0.43 and 2.00 $\mu_B$ when the lattice
constants of SrC were set to  4.70  and 5.67 \AA, respectively,
which are in agreement with the data in Fig.1 (a) of the
Letter\cite{dong11}. We can show that the mesh point number
$7\times7\times7$ in the work is much too few to converge the energy
and spin moment under pressure. As shown in Fig.\ref{fig:ek},
convergency of energy within 1 meV can only be achieved when the
k-points are larger than $20\times20\times20$ if the lattice
constant is 4.70 \AA. The convergency is faster if the lattice
constant is larger (e.g. 5.67 \AA) where the half metallic character
is present. This is mainly because of the metallic nature of the
Fermi surface which requires much more k-points to
converge\cite{martin}. The magnetic moment also changes
significantly with the increasing k-points and converges to zero
when the lattice constant is 4.70 \AA\ as shown in the figure with
hollow circles.
\par
\begin{figure}
\includegraphics[width=0.5\textwidth]{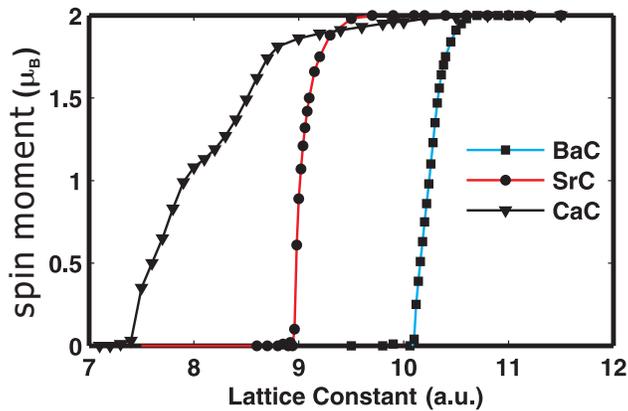}
\caption{The spin magnetic moment at different lattice constants of
the three compounds.}\label{fig:m-a}
\end{figure}

\begin{figure}
\includegraphics[width=0.5\textwidth]{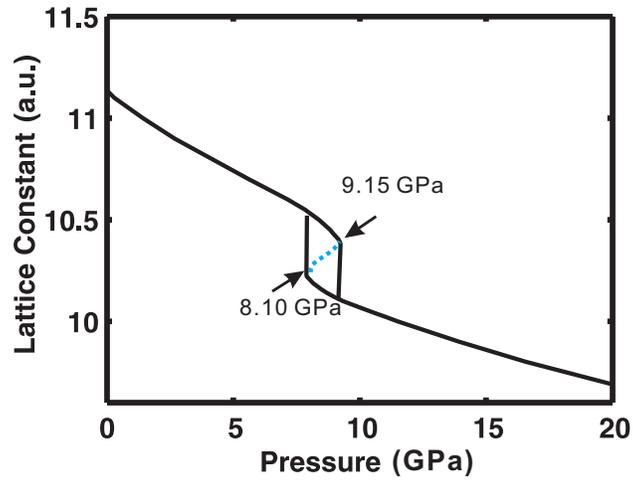}
\caption{The lattice constant vs. the hydrostatic pressure for BaC.
The dashed line shows the path of the unstable
state.}\label{fig:p-a-BaC}
\end{figure}

\begin{figure}
\includegraphics[width=0.5\textwidth]{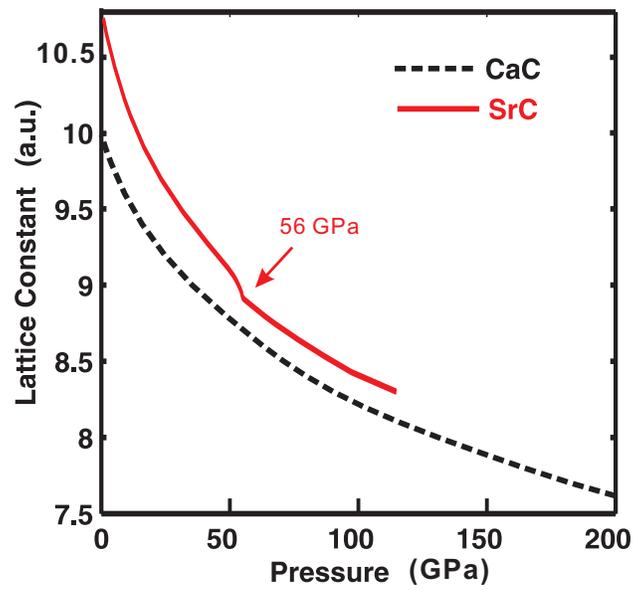}
\caption{The lattice constant vs. the hydrostatic pressure for SrC
(solid lines) and CaC (dashed lines).}\label{fig:p-a-SrC}
\end{figure}

 We recalculated the magnetic
moment at different lattice constants as shown in Fig.\ref{fig:m-a}
with the increased k-point number $32\times32\times32$. The spin
magnetic moment decreases continually to zero when the lattice
constant is decreased as shown in Fig.\ref{fig:m-a}. The decrease of
the spin moment in BaC and SrC is faster than in CaC. There is an
inflexion point when the lattice constant is 8.2 a.u. in the latter
compound. The equation of state of the three compounds was obtained
by differential of the energy with respect to the volume as shown in
Fig.\ref{fig:p-a-BaC}(BaC) and Fig.\ref{fig:p-a-SrC} (CaC and SrC).
There is hysteresis  in the pressure-lattice constant curves for BaC
which signatures a first order transition. This happens at
$\text{p}=9.14$ GPa when the pressure is increased and at pressure
$\text{p}=8.10$ GPa when the pressure is decreased. In the other two
compounds, the lattice constant changes gradually with the pressure.
However, a kink appears in SrC at pressure of 56 GPa.



\end{document}